# Solution Processability and Thermally Activated Delayed Fluorescence: Two Steps towards Low-Cost Organic Light Emitting Diodes

**Sahadev Somasundaram**

School of Chemical Engineering, The University of Queensland
St. Lucia, QLD 4072, Australia
d.somasundaram@uq.edu.au

## Abstract

Organic Light Emitting Diodes (OLED) have experienced vast attention in the scientific literature as the leading contender for the next generation of electronic lighting technology. However, while research continues to push the limit for OLED performance and efficiencies, minimal attention is paid to the processability and suitability for mass production. In this regard, Solution Processability (SP) and Thermally Activated Delayed Fluorescence (TADF) in OLEDs are two recent discoveries which have significantly improved the capacity for low-cost processing without compromise to their performance efficiencies. This meta-analysis is focused on these discoveries and their application to OLEDs. Recent peer-reviewed studies of OLEDs employing both SP and TADF are reviewed in order to highlight physical fundamentals and discover synergistic effects when incorporating both factors into a single OLED device. Additionally, some other recent trends targeting efficient manufacturing of OLEDs are identified, and logical future directions for interested researchers are discussed.

## Introduction

Organic light emitting diodes (OLED) are expected to serve as the next generation of technology for electronic lighting and display applications. While not yet sufficiently developed for broad commercialisation, OLEDs are beginning to show superior performance to their inorganic counterparts – in fact, several global electronics developers have opted for OLED-based displays in their product lines. The organic materials used in OLEDs carry added versatility in that they can maintain complete opto-electronic function when deposited onto flexible plastics or even fabric-type substrates. Furthermore, the underlying organic molecular structure allows for endless possibilities in tailoring production and performance efficiencies through unique molecular designs. Finally, recent research has focused on two inherent properties that dramatically improve production efficiency, which are only available to organic materials: Thermally Activated Delayed Fluorescence (TADF) and Solution Processability (SP).

The first notable demonstration of TADF by Chihaya Adachi and collaborators in 2012 [1] essentially quadrupled the possible performance efficiencies of conventional fluorescent OLEDs. By utilising a phenomena formerly termed "E-type delayed fluorescence" [2], the group discovered that purely organic materials can harness light emission from triplet-state excitons otherwise only accessible to materials containing expensive heavy metals. Thus, TADF materials not only improve the efficiency of conventional OLEDs but reduce the production cost by eliminating the need for expensive materials. Since discovery, most TADF studies have focused on achieving greater device performance through innovative molecular design. Over the previous few years, several reports have demonstrated achieving near 100% Internal Quantum Efficiency (IQE) [3,4], as well as External Quantum Efficiency (EQE) higher than the most promising phosphorescent OLEDs [5,6]. As a consequence, the production efficiency, or the ability to be manufactured at the commercial scale with minimal cost, receives considerably less attention. Studies of TADF-based materials and devices are widespread throughout the scientific literature; the mechanisms and significance of TADF are explained in numerous reviews, books and articles [1,7-14], and are thus not discussed here.

Solution processing is significantly cheaper than the typical vacuum-based deposition techniques, which require expensive equipment and a high resource cost per unit area of deposited material. Solution processing techniques also require less technical expertise for operation, are more suitable for large-scale depositions and are compatible with the existing roll-to-roll (R2R) processes used for mass production of thin-film semiconductors. Solution processing occurs from a solution of the deposition materials dissolved in a solvent, which is to be evaporated through the deposition process. It should be noted that while solubility is considered a major advantage of organic semiconductor materials, not all such materials are equally, or even sufficiently, soluble in suitable solvents; thus, solubility is a critical property when evaluating materials for solution processing. High solubilities are often associated with a greater presence of alkyl and alkoxyl moieties and a longer aliphatic chain [15]; this is typically seen in polymers. Higher solubility will i) result in a more uniform and homogeneous film morphology, removing the potential for charge trapping; and ii) allow a more uniform distribution of dopants in host:dopant complex films, resulting in a lower chance of concentration quenching [16] and aggregation-caused quenching [17,18].

Many industrial reports and popular science articles recognise that despite the exceptional performance of modern OLEDs, the cost involved with manufacturing is preventing them from dominating the electrical lighting and electronic display markets. It is believed that the key to these manufacturing issues lies in breakthroughs of the fundamental physics and chemistry of materials processing and OLED design, which is likely to be found in scientific literature and academic publishing (as opposed to grey literature).

Thus, the aim of this review is to:
- o Review all recent peer-reviewed papers examining OLED devices using a solution-processed and TADF EML;
- o Examine the concepts of TADF and solution processing in organic semiconductors, and identify any issues uniquely associated with OLED devices employing both phenomena, incl. synergistic effects; and
- o Highlight significant findings and identify a direction for future studies regarding solution processability, TADF and economic design of OLEDs.

Given the relatively low number of studies with exact relevance (i.e. development or discussion of OLED devices with both SP and TADF), articles are reviewed on a case-by-case basis and categorised with similar papers. Then, the findings are summarised and discussed with respect to the aims listed above. Articles not showing significant findings or discussions relevant to the advancement of SP TADF OLEDs or low-cost processing are not discussed here. Whereas other recent reviews on SP TADF OLEDs focus on novel molecular designs and structure-property relationships [19,20], this review retains focus on factors that contribute towards low cost manufacturing and the ability to be translated from research to commercialisation.

**Table 1.** Performance factors of solution processed, thermally activated delayed fluorescence OLEDs found in recent literature

| Reference | Photoluminescence Quantum Yield (PLQY) (%) | | Singlet-Triplet Splitting $\Delta E_{ST}$ (eV)[a] | Photolumiscence Wavelength, $\lambda_{PL}$ (nm) | | Maximum External Quantum Efficiency ($EQE_{max}$) (%) |
|---|---|---|---|---|---|---|
| | Solution | Film | | Solution | Film | |
| [21] | 26 | 26 | 0.1 | 550 | 540 | 9.3 |
| [22] | 94 | 52 | 0.03 | | | 2.4 |
| | 100 | 31 | 0.06 | | | 3.4 |
| | 94 | 8.5 | 0.06 | | | 1.5 |
| [23] | 92 | 52 | 0.13 | | | 6.0 |
| | 95 | 40 | 0.07 | | | 9.4 |
| | 96 | 45 | 0.08 | | | 9.5 |
| | 95 | 49 | 0.07 | | | 8.2 |
| [24] | 76 | | 0.07 | | | 16.1 |
| | 70 | | 0.09 | | | 16.0 |
| [25] | 34 | | 0.058 | 575 | | 17.5 |
| | 37 | | 0.048 | 608 | | 15.2 |
| [26] | 11 | 12 | 0.12 | 478 | | 1.0 |
| | 20 | 47 | 0.05 | 484 | | 8.9 |
| | 65 | 95 | 0.01 | 509 | | 13.9 |
| [27] | 55 | | 0.07 | 536 | 560 | |
| | 16 | | 0.31 | 429 | 424 | |
| [16] | 94 | | | | | 8.1 |
| | 67 | | | | | 8.2 |
| | 78 | | | | | 18.1 |
| [28] | 45 | | 0.18 | | | |
| | 98 | | 0.13 | | | |
| [29] | | | | | | 4.43 |
| [30] | | | | | | 12.9 |
| | | | | | | 12.4 |
| [24] | 44 | | 0.07 | | | |
| [31] | | | | | | 7.0 |
| | 68 | | 0.09 | 498 | 492 | 12.2 |
| | 48 | | 0.20 | 484 | 464 | 2.2 |
| [32] | | | | | | 9.4 |
| | | | | | | 9.9 |

| Ref | | | | | | |
|---|---|---|---|---|---|---|
| | | | | | | 15 |
| | | | | | | 16 |
| [26] | 11 | | 0.12 | 478 | | 1.0 |
| | 20 | | 0.05 | 484 | | 8.9 |
| | 65 | | 0.01 | 509 | | 13.9 |
| [33] | | | | 478 | | 1.2 |
| | | | | 492 | | 1.1 |
| | | | | 498 | | 4.0 |
| | | | | 506 | | 4.3 |
| [34] | 71 | | 0.16 | | | 18.7 |
| [35] | | | | | | ~24 |
| [36] | | | | | | 2.3 |
| | | | | | | 11.05 |
| [37] | 44 | | 0.18 | 590 | 553 | 1.2 |
| | 56 | | 0.17 | 594 | 545 | 5.5 |
| | 71 | | 0.20 | 591 | 541 | 11.8 |
| [38] | 42 | | 0.24 | | | 1.56 |
| | 61 | | 0.23 | 490 | 460 | 12.6 |
| [39] | 46 | 75 | 0.21 | 510 | 477 | 2.4 |
| | 2 | 39 | 0.05 | 608 | 522 | 3.7 |
| [40] | | | 0.24 | 439 | 445 | 23.8 |
| | 78 | | 0.22 | 452 | 475 | 23.5 |
| [41] | 12 | | 0.04 | | 549 | 8.1 |
| [42] | | 28 | 0.11 | | 572 | 5.8 |
| | | 52 | 0.05 | | 612 | 7.5 |
| [43] | | | | | | |
| | | | | | | |
| [44] | 3.8 | 25 | 0.15 | 407 | 473 | |
| | 65 | 65 | 0.08 | 483 | 505 | 4.3 |
| [45,46] | | 92 | 0.13 | 510 | 507 | 14.9 |
| [47] | 76 | 74 | 0.06 | 443 | | 17.8 |
| | 81 | 1.0 | 0.06 | 453 | | 20.0 |
| [48] | 77 | | 0.11 | 520 | | 12.0 |
| | 75 | | 0.15 | 499 | | 5.2 |
| [49] | 74 | | 0.04 | 606 | | 9.0 |
| [50] | 32 | | 0.28 | 502 | | 1.09 |
| | 58 | | 0.20 | 490 | | 6.5 |
| | 76 | | 0.20 | 487 | | 10.1 |
| [51] | 60 | 42 | 0.33 | 421 | | |
| | 64 | 58 | 0.25 | 438 | | |

| Ref | PLQY (%) | PLQY (%) | ΔE_ST (eV) | λ_PL (nm) | EQE (%) |
|---|---|---|---|---|---|
|  | 52 | 22 | 0.17 | 451 |  |
| [52] |  |  |  |  | 5.7 |
|  |  |  |  |  | 3.6 |
| [53] |  | 63 |  | 444 | 1.4 |
|  |  | 71 |  | 444 | 2.0 |
|  |  | 64 |  | 448 | 1.2 |
| [54] |  | 0.8 | 0.1 | 550 | 13.9 |
| [55] | 2.1 | 56 (powder) | 0.18 | 590 | 3.18 |
|  | 45 | 87 (powder) | 0.17 | 536 | 3.72 |
|  | 30 | 75 (powder) | 0.18 | 540 | 8.47 |
| [56] | 61 |  | 0.06 | 465 | 19.1 |
|  | 94 |  | 0.05 | 476 | 8.0 |
| [57] | 88 | 52 | 0.17 |  | 15.5 |
|  |  |  |  |  | 17.1 |
| [58] |  | 44 | 0.2 | 550 | 10.0 |
| [59] |  |  | 0.35 | 533 | 20.1 |
|  |  |  | 0.46 | 536 | 15.2 |
|  |  |  | 0.42 | 539 | 1.8 |
|  |  |  | 0.40 | 556 | 1.4 |
| [60] | 50 | 39 |  | 507 |  |
|  | 67 | 74 |  | 511 | 16.1 |
|  | 66 | 70 |  | 512 |  |
|  | 60 | 59 |  | 513 |  |

[a] Calculated via the singlet ($S_1$) and triplet ($T_1$) energies determined via onset of fluorescence and phosphorescence.

Table 1 presents a comparison of literature on OLED devices employing both SP and TADF within the emissive layer. To maintain concision, only primary performance figures are included to allow for general comparison of device performances across studies. Photoluminescence quantum yield, or PLQY (%), describes the ratio of photons emitted to photons absorbed during controlled photoluminescence in either solution or neat film states. Singlet-triplet splitting, or $\Delta E_{ST}$ (eV), represents the energetic difference between the lowest singlet ($S_1$) and triplet ($T_1$) electronic states, which can be conceptualised as the energetic gap electrons must overcome through Reverse Intersystem Crossing (RISC) and emit via the delayed fluorescence pathway. Photoluminescence wavelength, or $\lambda_{PL}$ (nm), describes the wavelength of emission during photoluminescence in the solution or neat film states, which is indicative of colour. External Quantum Efficiency, or EQE (%), describes the ratio of photons emitted to electrons injected during controlled electroluminescence, which reflects the overall device efficiency.

Detailed analyses of each study are included in the following sections.

## Polymers

The foremost benefit of polymers is their high solubility in organic solvents due to their long aliphatic chains. On the other hand, their large size and variability in molecular structure between production batches results in difficulties controlling molecular design factors that affect TADF emission performance. An example of such a design factor is the distances and torsions between donor and acceptor moieties; both must be separated in order to reduce overlap of the HOMO and LUMO (typically localised around the donor and acceptor, respectively), which is directly related to $\Delta E_{ST}$ [61].

In 2015, Nikolaenko, et al. [58] developed an SP TADF Emissive Layer (EML) made of a polymeric material. In this case, the donor and acceptor moieties served as distinct monomers in the greater polymer chain, and polymerisation amounts were controlled to vary charge transport properties and buffer the recombination zone. Other monomers were selectively added to control the electronic, optical and physico-chemical properties, such as solubility and viscosity, and a block-polymer structure was used to ensure tight control of polymer sequence. Interestingly, the experimental values of the containing devices showed a relatively low PLQY (in air) of 41% with estimated RISC yields of 83%, suggesting that light conversion, rather than triplet conversion, was limiting the efficiency of the device. Finally, the authors noted that the long-chain nature of the polymer EML may have resulted in horizontal alignment of the dipole moment of the EML polymer with the substrate, which is known to improve efficiency [62,63].

Nobuyasu, et al. [36] continued the foray into TADF polymers. Devices were made using two copolymers consisting of a known small molecule emitter. Both solution-processed polymer devices showed significantly lower efficiencies than vacuum-deposited small molecule devices and exhibited noticeable roll-off. Regardless, the authors noted that their polymer device efficiencies were comparable to those of the major polymer TADF effort by Nikolaenko, et al [58] the previous year. The low polymer performance was attributed to weak electronic properties as a result of unrefined solution-processing methods, without any further discussion given. The authors presented two primary challenges of TADF polymers: i) high molecular weight compounds cause high $\Delta E_{ST}$ and non-radiative internal conversion, effectively reducing the efficiency; and ii) Triplet-Triplet Annihilation (TTA) is frequent in polymers, creating an alternative non-radiative pathway for triplet excitons besides RISC. While the solubility of polymers makes them attractive for OLED devices, these issues need to be rectified before seeing commercial use.

Zhu, et al. [46] developed a conjugated polymer for use in a SP TADF OLED. In contrast to the typical alternating D-A backbone structure seen in most emissive polymers, acceptors were perpendicularly grafted to a backbone of donors to increase torsion, thus increasing HOMO-LUMO separation and reducing $\Delta E_{ST}$. Despite the unique electronic and optical properties of conjugated polymers for optoelectronics [64,65], the characteristic effects of conjugated polymers on the current device were not discussed.

The same polymer was used in an SP TADF OLED by Lin, et al. [45]. They observed high rates of triplet exciton quenching and attributed it to intra- and inter-molecular accumulation quenching between polymer chains. This accumulation quenching in TADF, which was recently studied, occurs via the close-range, electron-exchange Dexter energy transfer mechanism, as opposed to the long-range, dipole-dipole Forster mechanism [66]. To circumvent the accumulation quenching, they added an interfacial exciplex host to improve RISC rates and bolster the delayed fluorescence contribution. Indeed, quenching and roll-off were significantly reduced as evidenced by $L_{50}$ and $J_{50}$ (i.e. luminance and current density after EQE has reduced to 50% of maximum over prolonged operation), which were each increased nearly three-fold as compared to the pure polymer OLED.

A study by Xie, et al. [60] developed an SP TADF polymer using the same side-chain grafting strategy mentioned above and a TADF small molecule reported earlier [67,68] as an emissive repeating unit. The devices utilised a novel transfer mechanism where an additional TADF small molecule was included as an "assistant dopant", which mediates charge transfer between the host and TADF dopant and contributes towards a portion of the RISC processing of singlet excitons. Devices with optimised structure showed remarkable EQE of 16% with near 100% IQE. While the assistant dopant strategy is likely to improve device efficiencies, a three-component-EML is likely to require extra design considerations and, consequently, production costs for commercial production. Interestingly, the results also indicated that RISC rates are higher in polymers than small molecules.

Ren, et al. [59] developed a series of copolymer and homopolymer SP TADF OLEDs to illustrate the effects of polymer structure on TADF performance. In this case, emissive pendant moieties consisting of a phenothiazine donor and dibenzothiophene-dioxide acceptor were partitioned by inert styrene moieties in the greater polymer chain. This design was intended to boost emissive efficiency by reducing intermolecular interactions between pendant units. Indeed, the copolymers showed much lower $\Delta E_{ST}$ and significantly higher EQE, Current Efficiency (CE) and Power Efficiency (PE) than the homopolymer, and copolymers showed increasing performance along with increasing degree of copolymerisation (i.e. ratio of styrene units to emissive units). Partitioning the emissive units was believed to reduce quenching via internal conversion and triplet-triplet annihilation. However, no discussion was provided on the effect of molecular structure on solubility.

## Small Molecules

Small molecules are relatively low molecular weight (as compared to polymers) tightly-packaged molecules that form discrete particular neat films. As opposed to polymers, small molecules are not always soluble in ideal solvents, and thus present an extra complication if solution processing is desired. On the other hand, the tightly-packed, discrete particle film structure of small molecular neat films are more electrically conductive and thermally stable than polymers, and are thus more resistant to dissociation after formation of the neat film. A review of solution-processable small molecules for OLEDs can be found here [15].

In what might be the first of its kind, Cho, et al. [16] produced three separate solution processed TADF devices with $EQE_{max}$ values of up to 18.1%. One emitter – t4CzIPN – showed higher device efficiencies when fabricated via solution processing instead of vacuum deposition. In this case, it was believed that the t-butyl moiety improved solubility (0.4%) as compared to other designed emitters (0.1%) and resulted in a more stable film morphology. Further discussion purported that greater solubility resulted in a more uniform distribution in the final film which: i) subdues exciton quenching; and ii) minimises interstitials and defects, which lead to charge trapping – both of these factors will reduce overall quantum efficiency. Furthermore, when deposited in higher doping concentrations (3% and 5%), the resultant film morphology was noticeably more disrupted, which was believed to cause more charge trapping. Indeed, these devices showed lower current densities and quantum efficiencies. As noted earlier, the higher solubility of the t4CzIPN emitter might well be a result of the longer and more planar carbon backbone.

Cho et al. [47] developed highly efficient solution processed TADF emitters by modifying 4CzIPN, replacing either 1 or 2 CN groups with F atoms (3CzFCN and 4CzFCN). The rationale behind the design was to use the hydrophobic nature of F to counteract the hydrophilicity of CN and improve solubility in organic solvents. Indeed, the novel molecules showed higher solubility in toluene, and the solubility increased according to number of F moieties (1.5 wt% for 3CzFCN, 1.0 wt% for 4CzFCN and 0.1 wt% for 4CzIPN [69]). It is important to note that toluene is commonly required for TADF studies for its ability to prevent PL quenching of TADF molecules in solution. AFM revealed a much greater film morphology in 3CzFCN as a result of its higher dissolution in toluene. Furthermore, it was also believed that the weak electron withdrawing nature of the F moiety, relative to the parting CN, caused a slight bathochromic shift in the PL emission (443 nm and 453 nm for 3CzFPN and 4CzFPN, respectively, from 470 nm of 4CzIPN [69]). The study indicates that while functionalisation may be a simple yet effective method to improve solubility, PL or electrodynamic characteristics are likely to change – this might be particularly more pronounced in TADF molecules due to their delicate electronic nature.

The same group [34] later tried to improve the unstable emissive lifetimes of their previous efforts through a benzonitrile core substituted with 5 carbazole units (5CzCN). The aim of the design was to increase solubility by agglomerating as many donors as possible (while retaining the CN unit) in order to increase both the donor-acceptor dihedral torsion and the available surface area for solvent penetration. Indeed, the additional carbazole increased solubility in toluene from 0.1% for 4CzIPN, as per their previous study [69], to 0.5% for 5CzCN. However, it should be noted this figure is lower than the fluorinated benzonitriles of the group's previous effort mentioned above. Furthermore, while the lifetime was aimed to improve through extending conjugation of the benzonitrile acceptor and increase aromatic units through carbazole functionalisation, notable lifetime improvements were only detected in vacuum-deposited devices.

Kim, et al. [35] focused on improving production costs not only through soluble molecules, but through simplifying device structures. The aim was to create a device with no transport layers by using a "self-organised buffer hole injection layer" (Buf-HIL). 4CzIPN dopant and CBP host were used as the EML. The Buf-HIL consisted of PEDOT:PSS and a perfluoric acid copolymer (PFI), which resulted in a neat film with graded PFI concentration - the lower surface energy of PFI allows for natural self-organisation during spin coating. This gradient caused the work function to change across the Buf-HIL, allowing each end to closely match the electronic levels of its neighbouring layer and improve internal hole transport and reduce hole accumulation at the interface (i.e. work function of the bottom end was similar to the anode, and that of the top matched with the HOMO of the host material). Additionally, the Buf-HIL also improved efficiency by reducing contact between excitons and the lower energy levels of PEDOT:PSS, which has been reported to result in exciton quenching [70,71]. A polar aprotic solvent, THF, was selected to dissolve the highly polar 4CzIPN – indeed, solubilities were much higher in THF (>10%) than in toluene (<1%). THF not only improved film uniformity, it significantly reduced aggregation of 4CzIPN, resulting in a two-fold increase in PL intensity from a film spun from toluene as a result of aggregation-caused quenching. The authors found that solvents with higher dipole moments resulted in greater morphology and PL intensity. Ultimately, the study demonstrates a resourceful method to simultaneously improve both device efficiency and manufacturability.

Adachi's group [70] performed a comparative study of a series of 4CzIPN devices with either CBP or CPCB host, solution or vacuum processed layers and varying Electron Transport Layers (ETL). Interestingly, CBP devices showed no notable variation in $EQE_{max}$ between solution or vacuum processed devices, while those of CPCB showed up to a 31% decrease. Lifetimes were significantly lower in most devices when solution processed. However, solution processed CPCB devices had much improved lifetimes than solution processed CBP devices, which was attributed to a higher molecular weight, greater solubility and better film morphology.

Xie et al. [25] developed two small-molecule emitters that were both vacuum and solution deposited and compared. The solution deposited devices of each molecule showed comparable EQEs to their vacuum-deposited counterparts – this was even more notable considering that the vacuum deposited devices did not include hole transport layers.

Not long after the worldwide recognition of TADF, Chen, et al. [55] developed one of the first and most notable devices utilising both TADF and solution processability. The idea was to utilise the inherent narrow $\Delta E_{ST}$ found in transition metal complexes for TADF without the natural instability of rare phosphorescent transition metals like iridium or platinum. The same group [28] later developed a cuprous-complex small molecule which showed clear TADF in solution processed devices. The device was made by using a carbazole-based molecule, czpzpy, as both a ligand and host material for a separate emissive dopant, $[Cu(CH_3CN)_2(POP)]BF_4$ – both materials were simply spin deposited in tandem without prior preparation. The ligand was used in excess as an attempt to reduce dissociation of the emissive dopant and thus improve PLQY in the device.

Verma, et al. [54] developed a copper iodide complex for use in an inkjet-printed EML. A Copper-Iodide core was substituted with alkyl phosphites to improve solubility in nonpolar solvents suitable for inkjet deposition. The molecular design rationale was to include asymmetrically-substituted ligands in order to reduce the conformational possibilities, thereby increasing the entropy contribution and solubility. Solubilities in polar solvents such as ethanol were between 10 – 30 mg mL$^{-1}$ but were higher than 100 mg mL$^{-1}$ for non-polar solvents. The devices with a printed EML and evaporated organic layers showed an EQE$_{max}$ of 13.9%. With the broad solubility behaviour, the authors noted that the molecule was suitable for various types of high-perfomance inks and print heads – the molecule could be dissolved in solvents with tuneable viscosity or surface interactive behaviour.

Sun, et al. [53] developed solution processed TADF EMLs with thermally crosslinkable host and dopant. The desire for crosslinking EML materials was not made completely clear, but it was believed to contribute to a greater thermal stability and more homogenous film morphology. The device with a film of host:dopant mass ratio of 1:0.09 showed the best performance compared to those of 1:0.06 and 1:0.12, but the reason was not elaborated on.

Chen, et al. [56] developed a solution processed, non-doped, TADF emitter showing an EQE$_{max}$ of 19.1% - reportedly the highest as at publication (Oct 2017). This breakthrough was attributed to a dual-charge-transfer mechanism, in which charge was transferred both through conventional D-A transport as well as through-space electron transfer – in this case, electrons from the π-orbitals of donor and acceptor moieties in close proximity transfer through space [72,73]. Having an alternate pathway for charge transfer was believed to result in higher dipole moment and lower $\Delta E_{ST}$.

Huang, et al. [43] developed a TADF molecule with mechanochromic luminescence - colour variation depending on the solution with which the molecule was deposited. The molecule, Cz-AQ, emitted as a deep red at 680 nm when deposited through dichloroethane and yellow at 600 nm through a mixture of dichloroethane and ethanol (equivalent volume ratio). Crystals of the molecule were grown through different solutions and the crystallographies were studied to examine the luminescent behaviour. Between the two crystals, one showed stacking with an H-type aggregation along with *b* axis, leading to strong π-π interactions induced by significant orbital superimposition – this was believed to cause the red shift in emission of the deposited films. TGA and DSC also revealed transition of emission colour from red to yellow in the same crystal after thermal decomposition. Unfortunately, the specific solvatochromic effect on the molecule (i.e. the mechanism by which different solvents affect the molecular orientation and emission wavelength) was not discussed. It is also worth noting that due to the closely stacked molecular orientation, Cz-AQ showed aggregation-induced emission (AIE) properties – a mechanism in which luminescence is enhanced in aggregated states such as neat films [74].

## Dendrimers

Dendrimers serve as a new class of molecules with sizes in between those of small molecules and polymers. Dendrimers are identified by their tree-like structure, with "branch" molecules surrounding a single core in three dimensions. These branches then attach to subsequent branch molecules, thus giving the dendrimer a tree-like structure. Branch molecules are referred to as dendrons, while the number of branches in succession is termed "generation". Dendrimers are often designed using some existing small molecule as either dendrons or the core molecule. A typically high molecular weight ensures a good degree of solubility in common solvents, but also avoids forming inhomogeneous, impure or poorly distributed thin films that often occur in polymers with indefinite molecular weight [15]. While review articles of emissive dendrimers do exist [75,76], functional dendrimer OLED devices have only appeared in the literature within the last decade, and no relevant review articles have appeared since.

Albrecht, et al. [22] developed the first solution processable TADF OLED using a dendrimer-based emissive layer. The emissive dendrimers consisted of a triazine core unit attached to various generations of carbazole dendrons (up to 4 generations). Perhaps since Adachi's novel 4CzIPN molecule [1], Carbazole has become well-known moiety for functionalising TADF molecules because of its hole-transport velocity, high triplet energy, stable thermal resistance and its ability to improve solubility and resultant film morphology [31,77]. It was found that higher generation number resulted in worse performance in regard to $\Delta E_{ST}$ and PLQY (film and solution), but the highest EQE came from 3 generations. The increased intermolecular interactions resulting from higher generation numbers was believed to be the cause of lower PLQY. Regardless, each dendrimer showed clear TADF, indicating that dendrimers may well be the ideal solution to achieving both TADF and solution processability in future molecular designs. The authors also noted that carbazole-based dendrimers have been previously shown to become insoluble after photo-induced crosslinking [78] – this is hugely beneficial for solution processed devices, as it will allow the deposited EML to remain solid after contact with solution of neighbouring layers.

This orthogonal dendrimer and a device with fully solution-processed organic layers was later developed by the same group through the use of a similar triazine- and carbazole-based dendrimer [23]. Part of the strategy was to include terminating functional groups on the final generation of carbazole dendrons to increase initial solubility. Terminating groups included methyl, t-butyl, phenyl and dibenzocarbazole substituents. As expected, all funtionalised dendrimers showed good solubility in common solvents, such as chloroform, THF and toluene, and the dibenzocarbazole-functionalised dendrimer dissolved in hexane. However, no quantitative analysis nor qualitative speculation on the degree and change of solubility among variants was given. It was also shown that the functionalised dendrimers showed much better lifetimes than the non-functionalised – this was attributed to the electrochemical stability induced by the terminating groups, and the belief that the non-functionalised dendrimer might undergo electro-polymerisation in the powered device, which generated carrier traps.

The same group then continued their work and developed devices with t-butyl-substituted dendrimer as a dopant and a carbazole-dendrimer-based hosts [24]. The motivation was to select a host that could improve the PLQY of previously developed devices while maintaining orthogonality or resistance to dissolution in alcohol used to deposit ETLs. Hosts with generation numbers higher than 2 showed near-complete resistance to alcohol (likely due to higher molecular weight [79]), while that of 2$^{nd}$ generation was largely dissolved. Furthermore, a separate control experiment that even if the host is sufficiently resistant, alcohol can permeate the solid film of a host:dopant complex and dissolve the dopant, even if the dopant is intrinsically resistant to alcohol.

The most recent effort of Albrecht's group [52] involved a imparting AIE to a solution processable TADF dendrimer. Indeed, AIE was only noticed in dendrimers, as opposed to small molecules of similar composition, due to the intermolecular interactions occurring in dendritic structures – these interactions inhibit intramolecular rotations and vibrations, which create non-radiative pathways. As noted in previous studies, the dendrimers showed good resistance to alcohols, which allows for solution deposition of neighbouring ETLs in alcohol without degradation of the EML. While the EQEs of these dendrimers were comparatively lower than the authors' previous work, the synergy of AIE with TADF and solution processability serves as an innovative method to improve both performance and cost effectiveness of devices. Hu, et al. [39] and Huang, et al. [42] recently reported molecules with AIE, TADF and solution processability, albeit with much less notable results.

Li et al. [48] developed a new molecule by dendronising a DMAC-DPS core with up to two generations of t-butyl-substituted carbazole. Both dendrimers exhibited good solubility in dichloromethane, tetrahydrofuran, toluene and chlorobenzene, likely boosted by the t-butyl groups, but no quantitative solubility analysis was performed. High decomposition temperatures (471°C and 507°C for 1 generation and 2 generations, respectively) and post-annealing AFM revealed very high thermal stability and morphological homogeneity after spin coating for both dendrimers. However, different devices showed minor changes in efficiency according to varying annealing temperatures – EQEs for the 1-generation-dendrimer device showed fluctuations up to nearly 4 percentage points through annealing at 25°C to 160°C. Regardless, all devices showed relatively low roll-offs.

Ban, et al. [50] designed a dendrimer TADF emitter using alkyl, instead of π-conjugated or aromatic, chains to link dendrons to the emissive core. The rationale of this design was threefold: i) to improve solubility and film morphology by reducing molecular rigidity and steric hindrance; ii) to improve charge transport through non-conjugated linkers; and iii) to reduce emission quenching by using non-conjugated spacers to isolate the emissive core [80]. While the improvement to solubility and morphology was not verified, the alkyl-linked dendrimers indeed showed improved device performance, which was attributed to the effects predicted above. Electronic properties between dendrimers showed little variance (for example, relative standard deviation of $\Delta E_{ST}$ between dendrimers was only 13%) – this is interesting to note as steric hindrance and rigidity are known to be greatly influential on electronic states in TADF molecules. Thus, aliphatic linkers may well be one method to reconcile solution processability and TADF. The same group later witnessed similar effects through yellow- [37] and blue-emitting [38] dendrimers.

Ban et al. [57] later developed the first TADF OLED using fully solution processable organic layers. The dendrimer emitter, a cyanobenzene with two generations of carbazole dendrons (Cz-CzCN), was used as a non-doped (self-host) EML. The high molecular weight dendrimer was designed to provide resistance to alcohol solvents in order to avoid orthogonal dissolution when depositing the ETL. Thus, neat films of Cz-CzCN and a single-generation molecule (5CzCN) were subjected to absorption spectroscopy before and after rinsing in isopropyl alcohol - as expected, 5CzCN's absorption intensity reduced drastically, while that of Cz-CzCN only reduced by around 5%. Unfortunately, the mechanism behind the dendrimer's high alcohol resistance was not expounded on beyond the high molecular weight. Alkyl chains were used to link the 2$^{nd}$ generation dendrons, which were believed to enhance overall solubility and help maintain the electronic properties of the emissive core separate from dendrons. Furthermore, the encapsulating structure of dendrons was believed to limit exciton quenching via intermolecular interactions in the emissive core. Soon after, the same group [40] reported a new OLED device using similarly designed molecules (with 3 and 4 dendrons per generation instead of 5) as a host and guest EML, respectively. These devices were believed to use the same phenomena (i.e. inhibition of exciton quenching via molecular encapsulation and enhanced solubility via alkyl chain linkers) to achieve higher EQEs than the previous effort (EQE$_{max}$ of 23.8% and 23.5% for 3 dendrons and 4 dendrons, respectively). It should be noted that this appears to be the highest EQE reported for a solution processed TADF device to date.

# Summary

## Synergistic effects

There do not appear to be any beneficial synergistic effects between TADF and SP; in other words, the effects of TADF and SP do not combine to boost performance to levels unexpected by each factor individually. There are, however, notable detrimental effects - soluble materials have inherent properties that can affect the performance of TADFs and thus must be considered during design stages (and vice versa):

- o Deposition via solvent alters atomistic morphology and molecular orientation (as compared to solid crystalline structure), and thus effects the electron orbital configuration. Consequently, this has significant effects on the excitonic and electronic pathways, which will alter photo- and electroluminescent efficiencies;
- o Different types of solvents will alter emission characteristics, whether through alterations to the film structure, molecular orientation or both;
- o Both TADF contribution and solubility are currently precisely controlled by molecular structure – altering structure to tune one property is likely to affect the other. For example, increasing alkyl chain length to increase solubility may separate donor and acceptor moieties, disrupting the charge transfer mechanism or $\Delta E_{ST}$; and
- o Solution processed neat films are expected to have greater decay lifetimes [70]. This may disrupt the contributions to fluorescence from different sources (i.e. delayed or prompt).

It appears any breakthroughs in SP-TADF OLEDs will occur through trial and error and empirical observations of new molecular designs (as opposed to mechanistic analysis).

## Molecular Type

Polymers have higher solubilities as a result of the longer aliphatic chain. However, the electronic performance of TADF polymers is initially at a disadvantage as the length, complexity and variability of polymer structures introduces difficulties in tuning the structure-related properties of TADF (e.g. $\Delta E_{ST}$ and consequent IQE, which are related to D-A separation). On the other hand, small molecules are smaller and more precise in nature, allowing one to tune structure-related properties and performances through minor alterations to molecular structure or composition. Furthermore, they are also associated with simpler and more controllable synthesis methods, which is likely to reduce the costs of large-scale production. Finally, assuming a decent dissolution, small molecules generally produce well-ordered, neatly-packed thin films, resulting in greater electronic performance and lifetime.

Dendrimers are a great middle ground for low-cost design, capturing the solubility of polymers with the facile synthesis methods and the ability to precisely control molecular structure of small molecules. Dendrimers are suitable for TADF emission as the electronic structure and exciton transmission can be tuned through addition and modification of dendrons. Additionally, molecular weight can be largely varied by tuning the number and structure of dendrons; this allows one to selectively tune solubility and/or resistance to orthogonal solvents. Ken Albrecht's group has performed a number of studies into the characteristics of dendrimers and development of suitable dendrimer-based SP TADF OLEDs. Future work could focus on finding novel dendrimer and dendron structures for optimal performance along with suitable solvents and device structures.

# Future Research

## Solubility

While recognised as a critical factor for OLED design, few experimental studies actually discuss solubility or the consideration thereof before design. Since solubility is not easily quantified, many researchers simply rely on qualitative descriptors such as length of alkyl chains or molecular size to evaluate solubility. Glass transition temperature is one parameter that is associated with solubility [15,81]. While many of the studies reviewed herein do determine $T_g$ as part of experimental procedure, none have discussed it with regards to solubility. Instead, more effort should be given to quantify solubility and develop generic correlations between solubility and device stability or performance. In a review on SP small molecules for OLEDs, Duan, et al. [15] noted that the effect of solvent on solubility and film quality needs further investigation. The Solubility Parameter [82,83] attempts to numerically predict interactive dissolution via the intermolecular and diffusive forces. Derivations of the solubility parameter method have been used to study ink-based deposition of organic semiconductors in some theoretical studies [84-87], albeit with unpredictable results. Sanders, et al. [88] recently developed a methodical design process for OLED materials – determination of solubility parameters through computational molecular dynamics (MD) simulations was included in one step. Numerical correlations between solubility and resultant film properties would assist future researchers in SP OLED design.

## SP-electrodes

While solution processability is important in organic emissive materials, each device layer, including the electrode material, must be compatible with R2R processes in order to experience complete low-cost, large-scale manufacturing; this includes the electrode, which is typically metallic. Several articles have been devoted to SP metallic electrodes via deposition of metallic ink – a suspension of metallic nanoparticles in a colloidal system [89-93]. Unfortunately, as is common in bulk electrodes, this research focuses on the noble metals Au or Ag, which are not economically suitable for large-scale manufacture. As an alternative, Lee, et al. [94,95] have studied Al-based electrodes deposited via metallic ink and roll-to-roll process, and have recently reported a functional touch-sensitive panel with apparent resistance to oxidation. Cu, with its high inherent conductivity, may be another suitable candidate – as discussed earlier, Cu-based TADF molecules have already been solution processed. Widespread attention is given to carbon nanotube- or graphene- based cathodes, which have given promising electrical performances and are easily adapted to wet-deposition [89,96,97]. Polymer cathodes may be a viable alternative as they have been studied for use in batteries [98-103] and, to a lesser extent, nanoscale batteries [104] and solar cells [105].

Designers should also consider the solution processability of organic transport layers for fully SP OLEDs. However, transport layers have fewer design criteria than EMLs, and will not experience the same number of detrimental synergistic effects when depositing via solution as discussed earlier. Thus, there are fewer considerations when incorporating SP transport layers. Solvent orthogonality is perhaps the most significant consideration.

## Inkjet Deposition

The majority of SP-OLED studies use spin coating as the chosen method for deposition. While spin coating is indeed a wet process and allows one to analyse film structure, it does not necessarily translate into large-scale manufacture. Not only is the physical process of spin coating not applicable to the large scale, it cannot be incorporated into R2R processes. On the other hand, inkjet deposition is compatible and thus more suitable for commercial manufacture. In fact, electronic developers have recently begun using inkjet deposition in production lines or unveiled inkjet-deposited OLED prototypes. Thus, by using inkjet deposition in initial studies, researchers can ensure any developed materials are suitable for manufacture immediately and skip additional steps involved in translating their research to the market.

### Single-layer and non-doped devices

More efforts should be paid towards reducing production costs through single-layer devices and non-doped (i.e. no host) EMLs. Not only will this minimise production costs through reducing materials consumption, it will lessen the number of factors needing to be considered during design stages, thus simplifying the overall design procedure. As an example, doped EMLs require additional consideration of the entire electronic structure of the dopant along with that of the host. Similarly, in a fully-solution processed device, each layer needs a high level of orthogonal resistance to certain solvents to avoid dissolution when depositing neighbouring layers; single-layer devices will not require such consideration. In recent years, non-doped EMLs have made substantial progress in the literature, boasting greater stability and somewhat similar performance efficiency when compared to their doped counterparts [106]. On the other hand, single-layer OLEDs are considerably less common in research; the few notable recent studies of single-layer OLEDs in literature show significantly lower performance efficiencies (for example, see [107,108]). Light emitting electrochemical cells, which operate by a slightly different mechanism, are a perfect example of solid-state lighting through minimalistic single-layer design [109]. The Zysman-Colman group [110-113] have produced several light-emitting electrochemical cells using TADF EMLs, albeit at reduced efficiencies.

## Conflicts of interest

The author declares no conflicts of interest.